\documentclass[sigplan,screen]{acmart} % for camera ready

\usepackage{algorithm}
\usepackage{algorithmic}
\usepackage{todonotes,xspace}
\usepackage{graphicx}
\usepackage{caption}
\usepackage{subcaption}
\usepackage{amsmath}
\usepackage{mathtools}
\usepackage{csquotes}

\usepackage{fancyvrb}
\usepackage[frozencache,cachedir=minted-cache]{minted} % for ArXiv submission
\setminted[c]{
    frame=lines,
    framesep=2mm,
    baselinestretch=1.0,
    fontsize=\footnotesize, % a bit larger
    breaklines=true,
    xleftmargin=0em,
    xrightmargin=0em,
    autogobble,
    escapeinside=||, % 用于特殊注释处理
    showspaces=false,
    showtabs=false
}
\newcommand{\hlcomment}[1]{\textcolor{red!70!black}{/*@} \textcolor{red!70!black}{\textit{#1}} \textcolor{red!70!black}{*/}}
\newcommand{\spcomment}[1]{\textcolor{blue}{/*@} \textcolor{blue}{\textit{#1}} \textcolor{blue}{*/}}

\usepackage{framed}

\usepackage{adjustbox}
\usepackage{makecell}
\usepackage{tikz}
\usetikzlibrary{patterns,calc,arrows,shapes,snakes,automata,backgrounds,petri,positioning,decorations.pathmorphing,intersections}
\usepackage{tkz-euclide}
\tikzset{mynode/.style={draw,solid,circle,inner sep=1pt}}
\usepackage[customcolors]{hf-tikz}
\hfsetfillcolor{gray!10}
\hfsetbordercolor{none}
\usepackage{pgfplots}
\usepgfplotslibrary{units}  
\pgfplotsset{compat=1.17}
\usepackage{multirow}
\usepackage{pifont} 

\usepackage{hyperref}
\usepackage[
    type={CC},
    modifier={by-nc-sa},
    version={3.0},
]{doclicense}

\usepackage[capitalise]{cleveref}
\newcommand{\blue}[1]{\textcolor{blue}{#1}}

\newcommand{\ours}{\textsc{Preguss}\xspace}

\newcommand{\eva}{\textsc{Frama-C/Eva}\xspace}
\newcommand{\Wp}{\textsc{Frama-C/Wp}\xspace}

\newcommand{\mopsa}{\textsc{Mopsa}\xspace}

\def\triangleforqed{\hbox{$\lhd$}}
\makeatletter
\DeclareRobustCommand{\qedT}{%
	\ifmmode
	\eqno \def\@badmath{$$}%$$
	\let\eqno\relax \let\leqno\relax \let\veqno\relax
	\hbox{\triangleforqed}%
	\else
	\leavevmode\unskip\penalty9999 \hbox{}\nobreak\hfill
	\quad\hbox{\triangleforqed}%
	\fi
}
\makeatother

\newtheoremstyle{rmkstyle}{}{}{}{}{\bfseries}{.}{ }{\thmname{#1}\thmnote{ (#3)}}
\theoremstyle{rmkstyle}

\newcommand{\revision}[1]{#1}

\AtBeginDocument{%
  }

\setcopyright{acmlicensed}
\copyrightyear{2025}
\acmYear{2025}
\acmDOI{xx.xxxx/}
\acmConference[LMPL '25]{1st International Workshop on Language Models and Programming Languages}{October 15, 2025}{Singapore}
\acmISBN{979-8-4007-1248-7/2025/06}

\begin{document}

\title{\textsc{Preguss}: It Analyzes, It Specifies, It Verifies}
\subtitle{[Position Paper]}

\author{Zhongyi Wang}
\orcid{0009-0008-1986-6070}
\authornote{Both authors contributed equally to this research.}
\affiliation{%
  \institution{Zhejiang University}
  \city{Hangzhou}
  \country{China}}
\email{zhongyi.wang@zju.edu.cn}

\author{Tengjie Lin}
\orcid{0009-0002-5951-7314}
\authornotemark[1]
\affiliation{%
  \institution{Zhejiang University}
  \city{Hangzhou}
  \country{China}}
\email{tengjie.lin@zju.edu.cn}

\author{Mingshuai Chen}
\authornote{Corresponding author.}
\orcid{0000-0001-9663-7441}
\affiliation{%
  \institution{Zhejiang University}
  \city{Hangzhou}
  \country{China}}
\email{m.chen@zju.edu.cn}

\author{Mingqi Yang}
\orcid{0009-0004-5304-6763}
\affiliation{%
  \institution{Zhejiang University}
  \city{Hangzhou}
  \country{China}}
\email{mingqiyang@zju.edu.cn}

\author{Haokun Li}
\orcid{0000-0001-6411-9324}
\affiliation{%
  \institution{Peking University}
  \city{Beijing}
  \country{China}}
\email{ker@pm.me}

\author{Xiao Yi}
\orcid{0000-0002-4792-4433}
\affiliation{%
  \institution{The Chinese University of Hong Kong}
  \city{Hong Kong}
  \country{China}}
\email{yixiao5428@link.cuhk.edu.hk}

\author{Shengchao Qin}
\orcid{0000-0003-3028-8191}
\affiliation{%
  \institution{Xidian University}
  \city{Xi'an}
  \country{China}}
\email{shengchao.qin@gmail.com}

\author{Jianwei Yin}
% \authornotemark[2]
\orcid{0000-0003-4703-7348}
\affiliation{%
  \institution{Zhejiang University}
  \city{Hangzhou}
  \country{China}}
\email{zjuyjw@zju.edu.cn}

\renewcommand{\shortauthors}{Wang et al.}

\begin{abstract}
  Fully automated verification of large-scale software and hardware systems is arguably the holy grail of formal methods. Large language models (LLMs) have recently demonstrated their potential for enhancing the degree of automation in formal verification by, e.g., generating formal specifications as essential to deductive verification, yet exhibit poor scalability due to context-length limitations and, more importantly, the difficulty of inferring complex, interprocedural specifications. This paper outlines \textsc{Preguss} -- a modular, fine-grained framework for automating the generation and refinement of formal specifications. \textsc{Preguss} synergizes between static analysis and deductive verification by orchestrating two components: (i) potential runtime error (RTE)-guided construction and prioritization of verification units, and (ii) LLM-aided synthesis of interprocedural specifications at the unit level. We envisage that \textsc{Preguss} paves a compelling path towards the automated verification of large-scale programs.
\end{abstract}



% \received{28 May 2025}
% \received[revised]{1 June 2025}
% \received[accepted]{5 June 2025}

\maketitle

\section{Introduction}

Runtime errors (RTEs), e.g., division by zero, buffer/numeric overflows, and null pointer dereference, are a common cause of undefined behaviors (UBs) exhibited during the execution of C/C++ programs~\cite[Sect. 3.5.3]{c-standard}. These UBs can trigger catastrophic failures, rendering them critical considerations for safety-critical applications~\cite{ariane-5,buffer-overflow,resource-leak,runtime-error}.
%such as aerospace software~\cite{ariane-5} and medical devices~\cite{RTE-medical-devices,medical-devices}.\cmscomment{\atwzy Rephrase.} 
Consequently, in conventional program verification methodologies, establishing \emph{RTE-freeness} (i.e., conformance to the C standard specification) constitutes a necessary precondition for verifying \emph{functional correctness} (i.e., adherence to intended behaviors)~\cite{fm09-airbus}. State-of-the-art \emph{abstract interpretation}-based static analyzers, such as Astr{\'e}e~\cite{absint.astree}, \eva~\cite{eva-plugin}, and \mopsa~\cite{mopsa}, aim to reliably detect all potential UBs in C programs, thereby formally certifying the absence of RTEs. However, due to the inherent abstraction mechanism that soundly approximates concrete program semantics~\cite{cousotAbstractInterpretationUnified1977}, these tools often emit numerous false positives. Manually identifying such false alarms or tuning analyzer configurations for better accuracy remains notoriously difficult~\cite{parf}.

% \blue{Specification synthesis... Refer to AutoSpec, X509-parser, \#Spec to briefly introduce specification synthesis (and relationships with program verification) and why and how we can use ss for proving the absence of RTEs. (See the Section 2 Backgrounds.) Then give the main challenges of current ss methods.}

Program verification tools employing \emph{deductive verification}~\cite{deductive-verification} provide a rigorous methodology for ensuring critical program properties, including both RTE absence and functional correctness. The verification process typically involves two stages: (i) constructing \emph{specifications} to formalize intended program behaviors, and (ii) proving that the program adheres to the specifications. While modern verifiers such as \Wp~\cite{frama-c/wp} and Dafny~\cite{dafny} automate the latter stage, the former stage still relies heavily on human expertise. 

% The communities of formal methods, software engineering, programming languages, and artificial intelligence have witnessed a recent surge of interest in leveraging large language models (LLMs) for \emph{automated specification synthesis}, demonstrating substantial improvements over conventional techniques.

Recent studies~\cite{cav24-autospec,icse25-specgen,ASE24-Wu,ASE24-Pirzada,ICLR24-lemur} have explored large language models (LLMs) for \emph{automated specification synthesis}, demonstrating substantial improvements over conventional techniques. Nevertheless, these methods exhibit significant \emph{scalability limitations} when applied to large-scale programs. Current evaluations~\cite{cav24-autospec,ASE24-Wu,ASE24-Pirzada,ICLR24-lemur} remain confined to small-scale benchmarks (e.g., SV-COMP test suites~\cite{sv-comp24}, \revision{Code2Inv benchmark set~\cite{code2inv}, SyGuS competition~\cite{sygus-comp}}), each comprising individual or just a few functions. \revision{Although SpecGen~\cite[Sect. VI-A]{icse25-specgen} explores specification synthesis for substantial real-world Java programs, its focus remains restricted to specification \emph{verifiability} (i.e., syntactic and semantic correctness) rather than \emph{holistic program verification} -- emcompassing properties like RTE-freeness and functional correctness.} This limitation is primarily due to two constraints: First, the \emph{context-length limitation} of LLMs~\cite{LLM-context-length-limitation} prohibits them from processing large-scale programs as a whole; Second, verifying complex programs necessitates the synthesis of \emph{a diverse set of interprocedural specifications}, e.g., preconditions, postconditions, invariants, etc. The latter is beyond the capability of existing approaches: (i) some focus exclusively on specific categories (e.g., invariants~\cite{ICLR24-lemur, ASE24-Pirzada, ASE24-Wu}), and (ii) others generate \emph{function contracts} holistically without modeling intrinsic differences between preconditions and postconditions (as part of the contracts)~\cite{cav24-autospec,icse25-specgen}, which are critical for validating potential RTEs (as demonstrated in \cref{sec2:background-motivation}).

In response to the aforementioned challenges, we propose {\ours} -- an LLM-aided framework for synthesizing fine-grained formal specifications. {\ours} first employs a static analyzer to emit RTE assertions (signifying all possible RTEs), then constructs a queue of verification units (contexts of program slices and relevant assertions) according to the RTE assertions, and finally prompts an LLM to generate and refine interprocedural specifications for every verification unit. We envisage that {\ours} facilitates (i) the \emph{synergy between static analysis and deductive verification}: The RTE assertions reported by static analysis are used to either construct necessary specifications certifying RTE-freeness or locate root causes triggering RTEs; and (ii) \emph{modular synthesis of interprocedural specifications} and thereby a viable approach to the automated verification of large-scale programs.
\begin{figure}[t]
	\centering
		\begin{subfigure}[b]{0.48\linewidth}
			\centering
      \begin{minipage}{0.96\textwidth}
\begin{minted}{c}
#include <limits.h>
int abs(int x) {
  if (x < 0)
    |\hlcomment{assert overflow: -2147483647 <= x;}|
    return -x;
  else
    return x;
}
void main() {
  int a = abs(42);
  int b = abs(INT_MIN);
}
\end{minted}
    \end{minipage}
			\caption{analysis result of \texttt{abs.c}}
			\label{fig-1a:analysis-result}
		\end{subfigure}
		\hfil
		\begin{subfigure}[b]{0.5\linewidth}
			\centering
   \begin{minipage}{.97\textwidth}
\begin{minted}{c}
#include <limits.h>
|\spcomment{requires INT\_MIN < x;}|
int abs(int x) {
  if (x < 0)
    return -x;
  else
    return x;
}
void main() {
  int a = abs(42);
  |\hlcomment{preconditions of abs}|
  int b = abs(INT_MIN);
}
\end{minted}
\end{minipage}
			\caption{verification of specified \texttt{abs.c}}
			\label{fig-1b:verification-result}
		\end{subfigure}
	\caption{Identifying potential RTEs in a C program via the abstract interpretation-based analyzer {\eva}~\cite{eva-plugin} and the deductive verifier {\Wp}~\cite{frama-c/wp}.}
	\label{fig-1:example-analysis-verification}
\end{figure}

\section{Background and Motivation}
\label{sec2:background-motivation}

\subsection{Potential RTE-Guided Verification}

Abstract interpretation-based static analyzers~\cite{absint.astree, eva-plugin,mopsa} conduct sound value analysis of target programs to warn about potential RTEs.
%, such as signed integer overflows, invalid memory accesses, and null pointer dereferences.
These tools exhibit significant advantages in automatically scaling to large codebases. During the analysis, they embed \emph{RTE (guard) assertions} at program locations susceptible to undefined behaviors -- as exemplified by the assertion \textcolor{red!70!black}{\texttt{-2147483647 <= x}} in \cref{fig-1a:analysis-result}. The analyzers maintain \emph{abstract program states} -- e.g., overapproximating possible values $\{-3, 0, 3\}$ via an interval abstraction $[-3,3]$ -- and validate satisfiability of the RTE assertions against these abstract states. Unsatisfied assertions are subsequently reported as RTE alarms. \cref{fig-1a:analysis-result} illustrates the analysis result for the commonly used code snippet \texttt{abs.c} generated by {\eva}~\cite{eva-plugin}, which signifies a potential RTE of signed integer overflow expressed in the ANSI/ISO C Specification Language (ACSL)~\cite{acsl}.

Tracing the RTE in \cref{fig-1a:analysis-result} is a classic \emph{source-to-sink} problem~\cite{source-and-sink} in static (taint) analysis, namely, to identify that \texttt{x = INT\_MIN} (source) exclusively triggers the overflow RTE (sink) in the \texttt{abs} function. Although taint analysis tools can track data-flow paths from untrusted sources to sinks, validating these paths requires analysts to monitor how the data flows through the program and interacts with different components. This is a prohibitively challenging task in large-scale programs due to intricate call hierarchies and data dependencies. Deductive verification addresses this challenge through an advanced mechanism that propagates guard assertions from sinks upward along caller-callee chains, enabling violation checks at potential source locations. This approach facilitates either precise RTE source tracing or formal establishment of RTE absence~\cite{x509-parser,minimal-contract}.

Specifically, by augmenting functions with necessary specifications, e.g., the ACSL precondition \blue{\texttt{requires INT\_MIN < x}} in \cref{fig-1b:verification-result}, verifiers like \textsc{\Wp}~\cite{frama-c/wp} (based on weakest-precondition reasoning~\cite{weakest-precondition}) formally certify RTE-free execution under the specified preconditions. \emph{These preconditions subsequently serve as guard assertions at call sites of caller functions.} \cref{fig-1b:verification-result} demonstrates this mechanism: The argument \texttt{INT\_MIN} violates \texttt{abs}'s (weakest) precondition, thus leading the verifier to report a \emph{definitive} RTE at call site \texttt{abs(INT\_MIN)}; In other words, RTE-freeness can be guaranteed in case no caller of \texttt{abs} violates its precondition.

% This methodology -- termed \emph{potential RTEs guided verification} -- synthesizes abstract interpretation and deductive verification. Its core paradigm initiates from analyzer-reported RTE assertions to either (i) construct necessary specifications proving RTE freedom, or (ii) precisely localize root causes triggering RTEs.

This gives rise to our conceptual idea of \emph{potential RTE-guided verification}, which aims to synergize between static analysis and deductive verification. The core paradigm is to exploit analyzer-reported RTE assertions to either (i) construct necessary specifications certifying RTE-freeness, or (ii) locate root causes triggering RTEs. We will show in \cref{sec3:methodology} how this paradigm can be refined to a pipeline that facilitates automated verification of large-scale programs.

\begin{figure}[t]
	\centering
		\begin{subfigure}[b]{0.48\linewidth}
			\centering
      \begin{minipage}{0.96\textwidth}
\begin{minted}{c}
int id(int x) {return x;}

void one() {
  int x = id(1);
  |\hlcomment{assert division\_by\_0: x != 0;}|
  1/x;
}
void zero() {
  id(0);
}
void main() {
  one(); zero();
}
\end{minted}
    \end{minipage}
			\caption{analysis result of \texttt{id.c}}
			\label{fig-2a:analysis-result}
		\end{subfigure}
		\hfil
		\begin{subfigure}[b]{0.5\linewidth}
			\centering
   \begin{minipage}{.97\textwidth}
\begin{minted}{c}
|\spcomment{requires x != 0;}|
|\spcomment{\detokenize{ensures \result== x;}}|
int id(int x) {return x;}

void one() {
  int x = id(1); 1/x;
}
void zero() {
  |\hlcomment{preconditions of id}|
  id(0);
}
void main() {
  one(); zero();
}
\end{minted}
\end{minipage}
			\caption{verification of specified \texttt{id.c}}
			\label{fig-2b:verification-result}
		\end{subfigure}
	% \caption{Strong preconditions may introduce false positives in caller functions.}
    \caption{Demonstrating the dual role of interprocedural specifications: postconditions can eliminate false RTEs while over-constrained preconditions can induce false alarms.}
	\label{fig-2:dependencies-spec-property}
\end{figure}

\begin{figure*}
  \centering
  \includegraphics[width=0.98\linewidth]{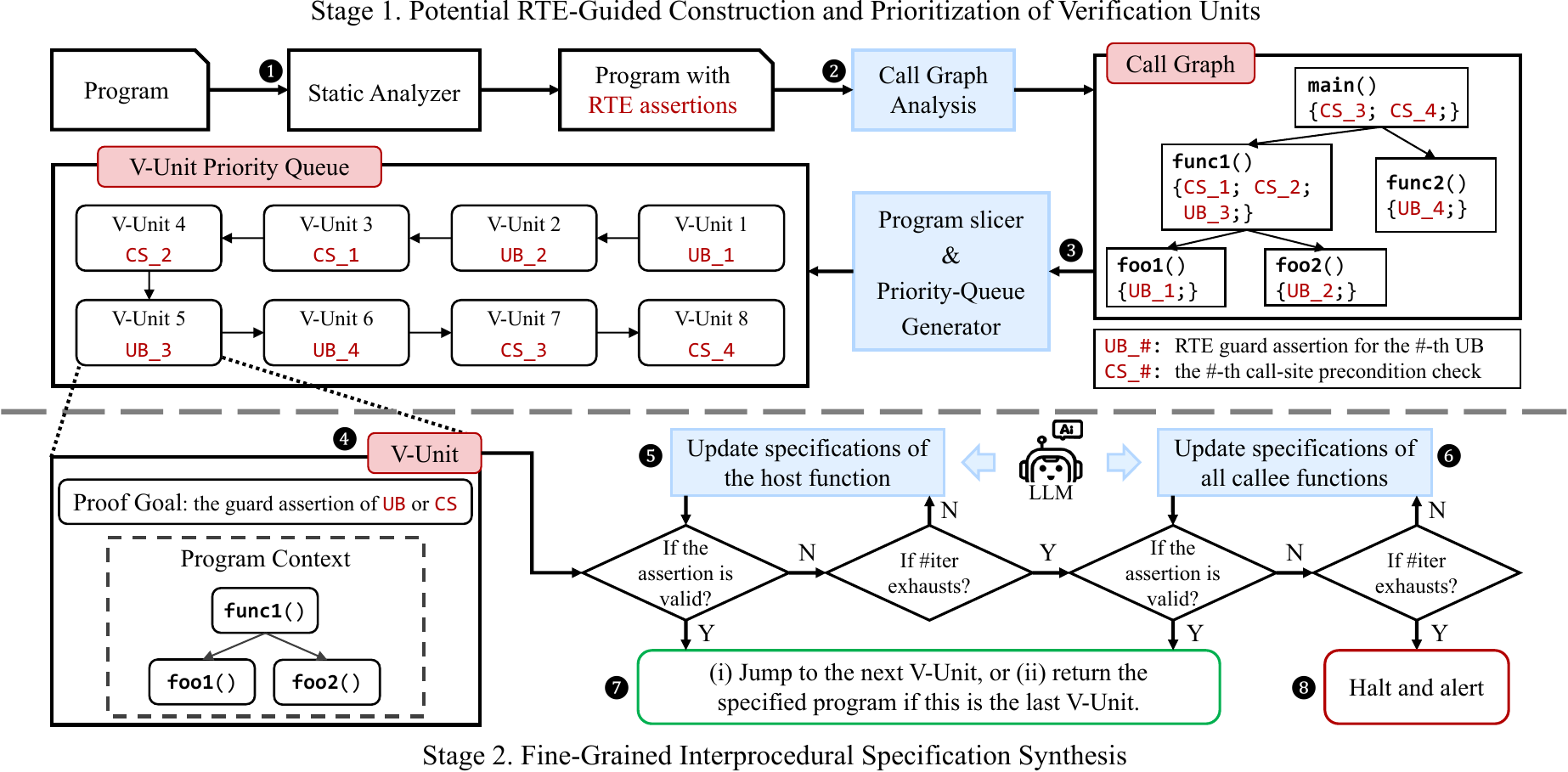}
  \caption{Architecture of the \ours framework.}
  \Description{...}
  \label{fig3:framework}
\end{figure*}

% \subsection{Strong Preconditions May Induce False Positives}
\subsection{Interprocedural Specifications}
\label{sec2-2:synergy-of-specifications}

The example in \cref{fig-1:example-analysis-verification} illustrates the simple case where an RTE assertion (\textcolor{red!70!black}{\texttt{-2147483647 <= x}}) can be validated through specifications (\blue{\texttt{requires INT\_MIN < x}}) \emph{localized to its host function} (\texttt{abs}). There are, however, common cases where \emph{interprocedural specifications} are necessary: As demonstrated by \cref{fig-2:dependencies-spec-property}, function \texttt{one} contains a potential division-by-zero UB flagged by assertion \textcolor{red!70!black}{\texttt{x != 0}} (\cref{fig-2a:analysis-result}), which is in fact a false alarm that can be eliminated solely through the postcondition \textcolor{blue}{\texttt{\detokenize{ensures \result== x}}} of function \texttt{id} (\cref{fig-2b:verification-result}). Concretely, the value of \texttt{x} is determined by argument \texttt{1} and the postcondition of \texttt{id} at call site \texttt{id(1)}, thereby guaranteeing assertion satisfaction irrespective of the precondition of its host function \texttt{one}. Both examples in \cref{fig-1:example-analysis-verification,fig-2:dependencies-spec-property} reveal a core forward verification principle: Validating control-flow downstream properties (e.g., RTE assertions in \cref{fig-1a:analysis-result,fig-2a:analysis-result}) requires correct upstream specifications (e.g., precondition in \cref{fig-1b:verification-result} or postcondition in \cref{fig-2b:verification-result}). This observation motivates specialized mechanisms for generating interprocedural specifications in programs with complex call hierarchies.

% Nevertheless, existing approaches~\cite{cav24-autospec,icse25-specgen} assume RTE-free targets for functional correctness verification and employ a simple interprocedural specification synthesis strategy -- prompting LLMs to generate function contracts holistically based on the identical program context. This strategy may induce false positives when validating RTE-freeness, as exemplified in \cref{fig-2:dependencies-spec-property}. LLMs sometimes can be mislead by part of the context, such as the statement \texttt{1/x} susceptible to division-by-zero UB and the assertion \textcolor{red!70!black}{\texttt{x != 0}} in function \texttt{one}, and thus yield excessively strong preconditions \textcolor{blue}{\texttt{requires x != 0}}, which  triggers a spurious alarm at call site \texttt{id(0)} in \texttt{zero}. Someone may notice that by eliminating the precondition while retaining the postcondition allows proving full RTE-freeness of the entire program. However, potential RTEs guided verification \textit{repositions preconditions as RTE guards}, and thus violated preconditions may denote genuine errors (e.g., \cref{fig-1b:verification-result}). Therefore, \emph{the special mechanism for interprocedural specification synthesis should avoid generating over-constrained preconditions}.

Nevertheless, existing approaches~\cite{cav24-autospec,icse25-specgen} mostly assume RTE-freeness of a target program for verifying functional correctness, employing simplistic interprocedural-specification synthesis strategies that ask LLMs to generate \emph{holistic} function contracts across the entire program context in a uniform, monolithic manner. This methodology risks inducing false positives during RTE-freeness validation, as illustrated by \cref{fig-2:dependencies-spec-property}. An LLM may be misled by the context \enquote{\textcolor{red!70!black}{\texttt{x != 0}} and \texttt{1/x}} in \texttt{one} and thus forge an \emph{over-constrained} precondition \textcolor{blue}{\texttt{requires x != 0}} for \texttt{id}, which triggers spurious alarms at call sites like \texttt{id(0)} in function \texttt{zero}. Although discarding over-constrained preconditions while retaining postconditions may resolve immediate false alarms (as is the case for \cref{fig-2b:verification-result}), determining whether a precondition is over-constrained is per se a nontrivial task (cf.\ \cref{fig-1b:verification-result} vs.\ \cref{fig-2b:verification-result}). For \cref{fig-1b:verification-result}, simply discarding the precondition compromises soundness (as true RTEs are missed) and thus
%introduces significant verification risks. In our potential RTE-guided verification framework, preconditions are fundamentally repositioned as RTE guards, and therefore, precondition violations inherently denote genuine errors. Arbitrary removal of these guard conditions would 
undermines verification integrity. Therefore, interprocedural-specification synthesis mechanisms should make efforts to prevent the generation of over-constrained preconditions.

\section{Methodology}
\label{sec3:methodology}

This section presents the design principles behind {\ours} -- our framework for Potential Runtime Error-GUided Specification Synthesis. As depicted in \cref{fig3:framework}, {\ours} is comprised of two synergistic components: (i) \emph{Potential RTE-guided construction and prioritization of verification units}: a divide-and-conquer strategy~\cite{divide-and-conquer} for decomposing the monolithic RTE-freeness verification into prioritized units; (ii) \emph{Fine-grained interprocedural specification synthesis}: a tactic for inferring necessary specifications via LLMs along caller-callee chains to validate target assertions per verification unit. Below, we show how these two components cooperate to facilitate {\ours}'s scalability to large-scale programs.

\subsection{Decomposing the Monolithic Verification}
\label{sec3-1:construction-prioritization}

% To overcome \emph{context length limitations} of LLMs, we decompose whole-program RTE-freeness verification into sequential validation of individual potential RTEs with necessary program context. Specifically, we first employ static analysis to generate RTE assertions for all possible UBs in the target program (\ding{182}). We then construct the program's call graph while recording call sites (\ding{183}), thereby obtaining all RTE guard assertions. Notably, although UB assertions are generated during initial analysis, the actual RTE guard assertions for call sites initially default to the tautological \texttt{true} and will be subsequently generated or updated by LLMs following the strategy detailed in \cref{sec3-2:interprocedural-spec-syn}. For each assertion, we construct a Verification Unit (V-Unit) containing both the assertion and necessary program slices. These V-Units are then prioritized through post-order traversal of the call graph (\ding{184}), adhering to bottom-up verification principles that progress from leaf functions toward the root~\cite{cav24-autospec}.

To address the context-length limitation of LLMs~\cite{LLM-context-length-limitation}, we decompose the monolithic RTE-freeness verification problem into \emph{a sequence of subproblems for validating individual potential RTEs associated with necessary program contexts}. As shown in \cref{fig3:framework}, we first employ static analysis to generate RTE assertions for all possible UBs in the target program (\ding{182}). We then construct the program's call graph while recording all the call sites (\ding{183}), yielding a complete set of RTE guard assertions. These assertions are classified into two distinct categories: (i) UB assertions generated during the initial analysis, and (ii) call-site preconditions (initially defaulting to tautological \texttt{true}) to be synthesized/updated by LLMs per \cref{sec3-2:interprocedural-spec-syn}. For each assertion, we construct a \emph{verification unit} (V-Unit) containing both the assertion and its necessary contextual program slices. These V-Units are prioritized in a queue (\ding{184}) exactly matching the sequence of their corresponding RTE assertions produced by a post-order traversal of the call graph. This deliberate ordering implements the bottom-up verification mechanism by progressing systematically from leaf functions toward the root function \cite{cav24-autospec}.

\ours leverages the \emph{modularization principle of deductive verification}~\cite[Chap.~9]{deductive-verification}, wherein functions are verified in isolation using their contracts. Rather than feeding LLMs with the entire program in an end-to-end manner, validating individual RTEs requires a constrained context comprising two components: the guard assertion and a two-layer program slice (encompassing both the assertion's host function and its callees), collectively encapsulated within a V-Unit (e.g., \ding{185}). \emph{This design ensures that the context does not expand significantly during the bottom-up verification progression, thereby enabling scalability to large-scale programs.} %The process for generating fine-grained interprocedural specifications within V-Units is elaborated in \cref{sec3-2:interprocedural-spec-syn}.

\subsection{Generating Interprocedural Specifications}
\label{sec3-2:interprocedural-spec-syn}

As motivated in \cref{sec2:background-motivation}, {\ours} implements a fine-grained strategy for synthesizing interprocedural specifications. This approach fundamentally differs from existing methods by avoiding the holistic function-contract generation from uniform contexts. {\ours} processes each V-Unit through two distinct phases: specification generation for the assertion's host function (\ding{186}), followed by specification generation for callee functions (\ding{187}). Both phases employ an iterative refinement strategy where (i) an LLM generates candidate specifications, (ii) a verifier validates the target assertion using these specifications, and (iii) an LLM refines the (possibly) failed candidates based on the verifier's feedback. This procedure terminates until either the validation succeeds or the iteration limit (\#iter) is exhausted. In the initial phase (\ding{186}), the LLM is prompted to infer preconditions and auxiliary specifications (e.g., loop contracts) through backward reasoning from the guard assertion -- as informally analogous to the process of weakest-precondition reasoning~\cite{weakest-precondition}. %\blue{Postcondition generation is intentionally \emph{excluded} in this phase since such specifications contribute minimally to validating assertions within host functions.}\wzycomment{Overstatement.}
~When the assertion depends on callees' return values (recall \cref{fig-2a:analysis-result}), the verification requires callees' postconditions, thus triggering the transition to the subsequent phase (\ding{187}). Crucially, to prevent over-constrained preconditions (as exemplified in \cref{fig-2b:verification-result}), the callees' precondition synthesis is expressly prohibited during this stage. Upon successful verification of the assertion alongside all generated specifications (excluding host preconditions), \ours either proceeds to the next V-Unit or returns the specified program if the priority queue becomes empty (\ding{188}). Unverifiable assertions trigger immediate termination with high-risk RTE alerts (\ding{189}).

% In conclusion, inspired by the forward verification principle illustrated in \cref{sec2-2:synergy-of-specifications}, \ours leverages the limited but sufficient context provided in each V-Unit and takes a two-stage strategy to construct different interprocedural specifications. This fine-grained strategy aims to avoid generating over-constrained preconditions, thereby making it possible to verify RTE-freeness of large programs.

% {\ours} directly operationalizes \cref{sec2-2:synergy-of-specifications}'s insights: by leveraging the V-Unit's constrained context and implementing a two-stage interprocedural-specification-synthesis strategy -- mirroring the control-flow upstream/downstream dependency principle -- \ours systematically avoids generating over-constrained preconditions while enabling scalable RTE-freeness verification for large programs.

{\ours} directly operationalizes our insights presented in \cref{sec2-2:synergy-of-specifications}: By leveraging the V-Unit's constrained context and implementing a two-stage interprocedural-specification synthesis strategy (thereby mirroring the forward verification principle), \ours systematically avoids generating over-constrained preconditions while enabling automated, scalable RTE-freeness verification for large-scale programs.

\section{Conclusion and Future Directions}
\label{sec4:conclusion-future-directions}

We have conceived {\ours} -- a modular, fine-grained framework for inferring formal specifications that synergize between static analysis and deductive verification. We envisage that {\ours} offers a promising paradigm towards the automated verification of large-scale programs (subject to an extensive experimental evaluation in future work).

While initially devised to certify RTE-freeness and identify genuine RTEs, the framework can be extended to cater for the verification of other vulnerability classes and functional correctness. The key is to substitute RTE-assertion generation (via static analysis) with tailored formal annotations for the target properties, including ones with complex data structures. The subsequent stages, commencing from \ding{183} in \cref{fig3:framework}, remain fully generic to handle these annotations.

\revision{Although \ours presents a promising paradigm for automated large-scale program verification, establishing a formal proof of its \emph{soundness} remains non-trivial -- It requires a rigorous justification that verified properties hold under all execution paths. This necessitates foundational work in two directions: (i) precise formalization of target verification properties, and (ii) rigorous proof establishing semantic equivalence between synthesized specifications and program behaviors. This foundational work must bridge the gap between practical LLM-based synthesis and formal method guarantees. Additionally, language-specific complexities pose significant soundness threats. The pervasive use of function pointers in C programs may induce incomplete call graphs during static analysis, potentially causing undetected function dependencies, verification context omissions, and unsound RTE guard propagation. Ensuring call graph reliability -- through advanced pointer analysis techniques like flow-sensitive analysis -- becomes imperative for any robust implementation.}

Conceived for scalability in programs with lengthy chains of function calls, \ours exhibits limitations when applied to: (i) Mutually recursive functions, which induce loopy structures in the call graph disrupting the bottom-up verification mechanism, causing potential failures of the interprocedural specification synthesis; (ii) Star-structured call graphs (characterized by few callers invoking numerous callees), where V-Units containing an inflated context of such functions may exceed the LLM's context limit. For the latter, a potential solution is to employ program slicing to extract the minimal subset of statements exhibiting dependencies with the target assertion, thus omitting nonessential callees.

\revision{Given LLMs' susceptibility to \emph{hallucination}~\cite{hallucination}, they frequently generate specifications exhibiting critical flaws -- ranging from \emph{syntactically illegal} constructs that violate specification language grammars to \emph{semantically unsatisfiable} predicates that contradict actual program behaviors. While verifiers provide rich feedback (e.g., proof obligations and diagnostic logs) on such errors, translating these formal outputs into effective LLM refinement prompts presents a significant research challenge due to the formal-to-informal semantic gap (cf.\ \cite{acl25}). Future work should therefore focus on developing principled mechanisms to bridge this gap.}

\begin{acks}
    %% acks environment is optional
    %% contents suppressed with 'anonymous'
    %% Commands \grantsponsor{<sponsorID>}{<name>}{<url>} and
    %% \grantnum[<url>]{<sponsorID>}{<number>} should be used to
    %% acknowledge financial support and will be used by metadata
    %% extraction tools.
    %   This material is based upon work supported by the
    %   \grantsponsor{GS100000001}{National Science
        %     Foundation}{http://dx.doi.org/10.13039/100000001} under Grant
    %   No.~\grantnum{GS100000001}{nnnnnnn} and Grant
    %   No.~\grantnum{GS100000001}{mmmmmmm}.  Any opinions, findings, and
    %   conclusions or recommendations expressed in this material are those
    %   of the author and do not necessarily reflect the views of the
    %   National Science Foundation.
    %
    \revision{This work has been partially funded by the ZJNSF Major Program (No.\ LD24F020013), by the CCF-Huawei Populus Grove Fund (No.\ CCF-HuaweiSY202503), by the Open Fund of the High-Reliability Embedded Software Engineering Technology Laboratory (No.\ LHCESET202502), and by the Huawei Technical Collaboration Project (No.\ TC20250422031).}
\end{acks}

% \cmscommentinline{Page limit: 2-4 pages excluding references.}

\bibliographystyle{ACM-Reference-Format}
\bibliography{references}

\end{document}